\documentstyle[11pt,paspconf,epsf]{article}
\input{psfig.tex}

\begin{document}

\title{Blazars: recent developments}

\author{Gabriele Ghisellini}
\affil{Osservatorio Astronomico di Brera, V. Bianchi 46, I-23807 Merate Italy}

\begin{abstract}
Recent observational and theoretical results on blazars
are presented and discussed. 
We are beginning to understand the rich phenomenology of blazars, and we are 
finding trends which will hopefully lead us to unveil the physics of
these extreme sources.
Limits and constraints on how jets originate can be derived
by their intense and variable $\gamma$--ray emission. 
We are now starting to study their variability behavior simultaneously 
at different frequencies, to find delays and lags, to put constraints
on emission models and source geometry.
We observe trends in the overall spectral energy distribution.
Also the problems of the overall energetics of jets
and of their matter content are now being tackled.
\end{abstract}


\section{Introduction}
EGRET, onboard CGRO, and ground based Cerenkov telescopes detected 
intense and variable $\gamma$--ray emission from blazars
(von Montigny et al. 1995; Thompson et al. 1995, 1996;
Weekes et al. 1996; Petry et al. 1996).
This discovery has allowed a much deeper understanding of 
their emission processes and, more generally, of the physics
of relativistic jets.
We can at last construct the entire spectral energy distribution (SED)
of blazars, and know that, during flares, the $\gamma$--ray power output
can greatly exceed what is emitted in the rest of the electromagnetic spectrum.
This discovery has revived blazar research and further stressed
the need for strictly simultaneous observations in different bands,
since fluxes emitted at different frequencies may be produced by the same 
process or within the same emitting region.

\section{A few things that $\gamma$--ray tell us}

The very fact the we observe a strong and variable $\gamma$--ray flux
from blazars bears very important consequences.

The first implication is that the $\gamma$--rays we see are not 
absorbed by the $\gamma$--$\gamma \to e^+e^-$ process, the most effective 
$\gamma$--ray absorption process in the blazar disk/jet environment.
This means that the hard X--ray and the $\gamma$--ray radiation
must be beamed:
in this way the photon density in the comoving frame is reduced, and
the source can be transparent to the high energy emission (see Dondi
\& Ghisellini 1995 for the derivation of the beaming factor
for a sample of EGRET sources).

The second implications is that there cannot be, even externally to the 
$\gamma$-ray production site, any strong source of X--rays,
which are target photons in the $\gamma$--$\gamma \to e^+e^-$ process.
Therefore not only the emitting plasma must be in relativistic
bulk motion, but it also has to be far away from any strong X--ray source,
such as the hot corona sandwiching the accretion disk.
This must be true not only for the region emitting
the $\gamma$--rays reaching us, but throughout the entire jet.
Let us consider, in fact, the inner part of the jet (closer to the black hole), 
and assume that it emits a significant fraction
of the total power in $\gamma$--rays and that they get absorbed, originating 
$e^\pm$ pairs, as in the Blandford \& Levinson (1995) model.
These pairs are born relativistic, and reprocess, through inverse Compton
scattering, the UV radiation produced by the disk.
The inverse Compton and the photon--photon pair creation processes
have roughly the same
cross section: the relative importance of the two processes is determined
by the relative energy densities in X--rays vs IR--UV photons.
The result is that those pairs reprocess the power originally in $\gamma$--rays
into X--rays.
We should then observe roughly equal X--ray and $\gamma$--ray luminosities
(the latter coming from more extended, optically thin parts of the jet).
This is not observed.
Most of the radiative power must therefore be produced far away
(hundreds of Schwarzchild radii) from the black hole.
{\it There must be some energy transport mechanism which is dissipationless 
up to these distances} (Ghisellini \& Madau, 1996).
Candidates are Poynting flux and collimated relativistic protons.

\section{Many kinds of variability behaviors}

Fast and large amplitude variability is a defining property of blazars
(see the review by Ulrich, Maraschi \& Urry, 1997).
However, not all blazars share the same variability behaviour, and the `old' 
rule of more extreme variability for larger frequencies is not always true.
Even in this research area the discovery of GeV--TeV emission of blazars 
has been crucial in stimulating many (but not yet enough) observational 
campaigns aiming to follow  single sources from the radio to the TeV 
band (see the review by Wagner 1997 and references therein).
Fig. 1 tries a first naif classification of the variability we see.
Variability in different bands is correlated, suggesting that a single
populations of electrons is responsible for producing the two broad peaks
of emission by the synchrotron and inverse Compton processes.
Time lags are starting to be observed, and even if these results are still
somewhat puzzling (e.g. the different variability behavior of PKS 2155--304 
in two different campaigns, see Ulrich, Maraschi \& Urry, 1997 and
references therein), this is one of the most powerful way
to study the acceleration mechanism, and to put strong
constraints on emission models (Kirk, 1997 and references therein;
Tavecchio, Maraschi \& Ghisellini, 1998).
In Fig. 2 some SEDs of specific blazars are shown, to illustrate concrete 
examples of the tentative variability classification sketched in Fig. 1.

\begin{figure}
\vskip -2 true cm
\psfig{file=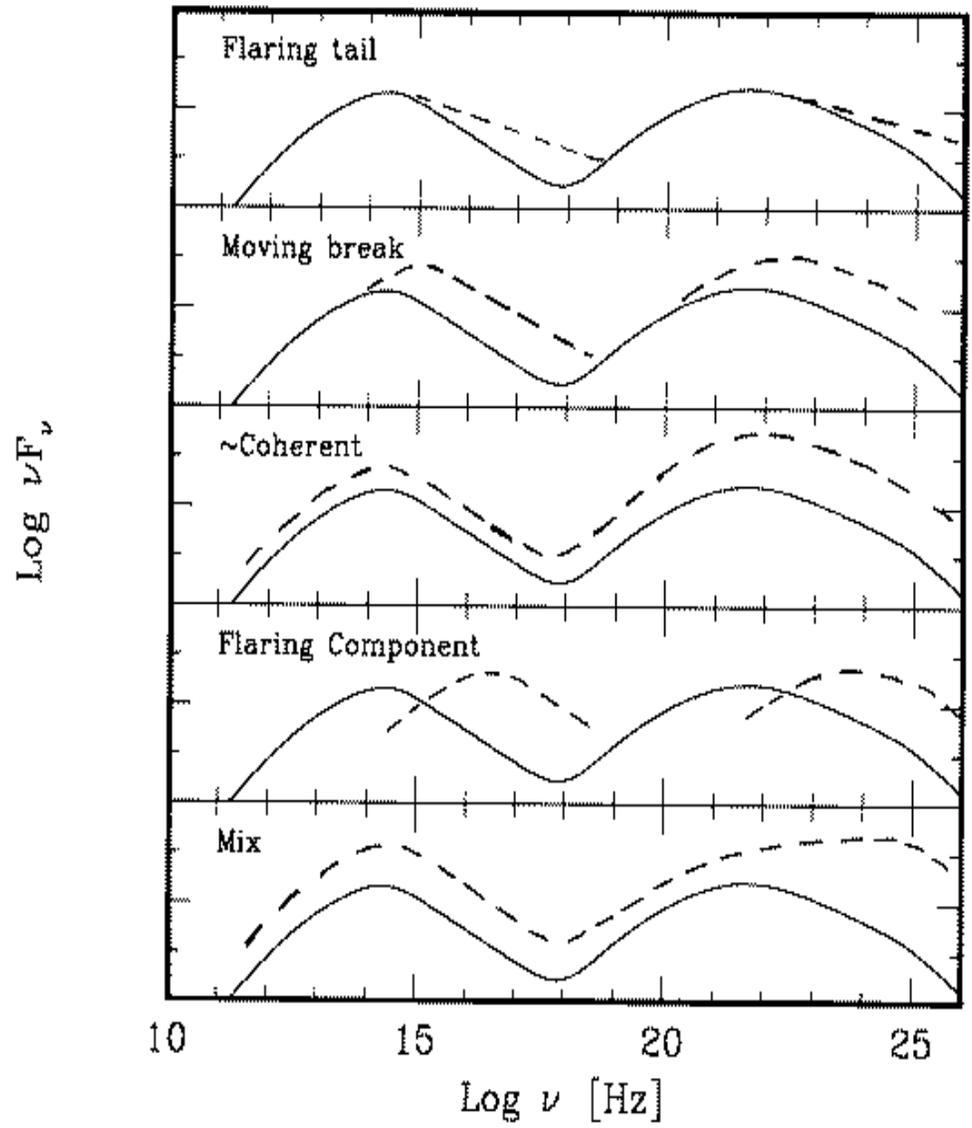,width=15truecm,height=21truecm}
\vskip -1.5 true cm
\caption[h]{Schematic examples of the observed overall variability behavior of
blazars.}
\end{figure}

\begin{figure}
\vskip -2 true cm
\psfig{file=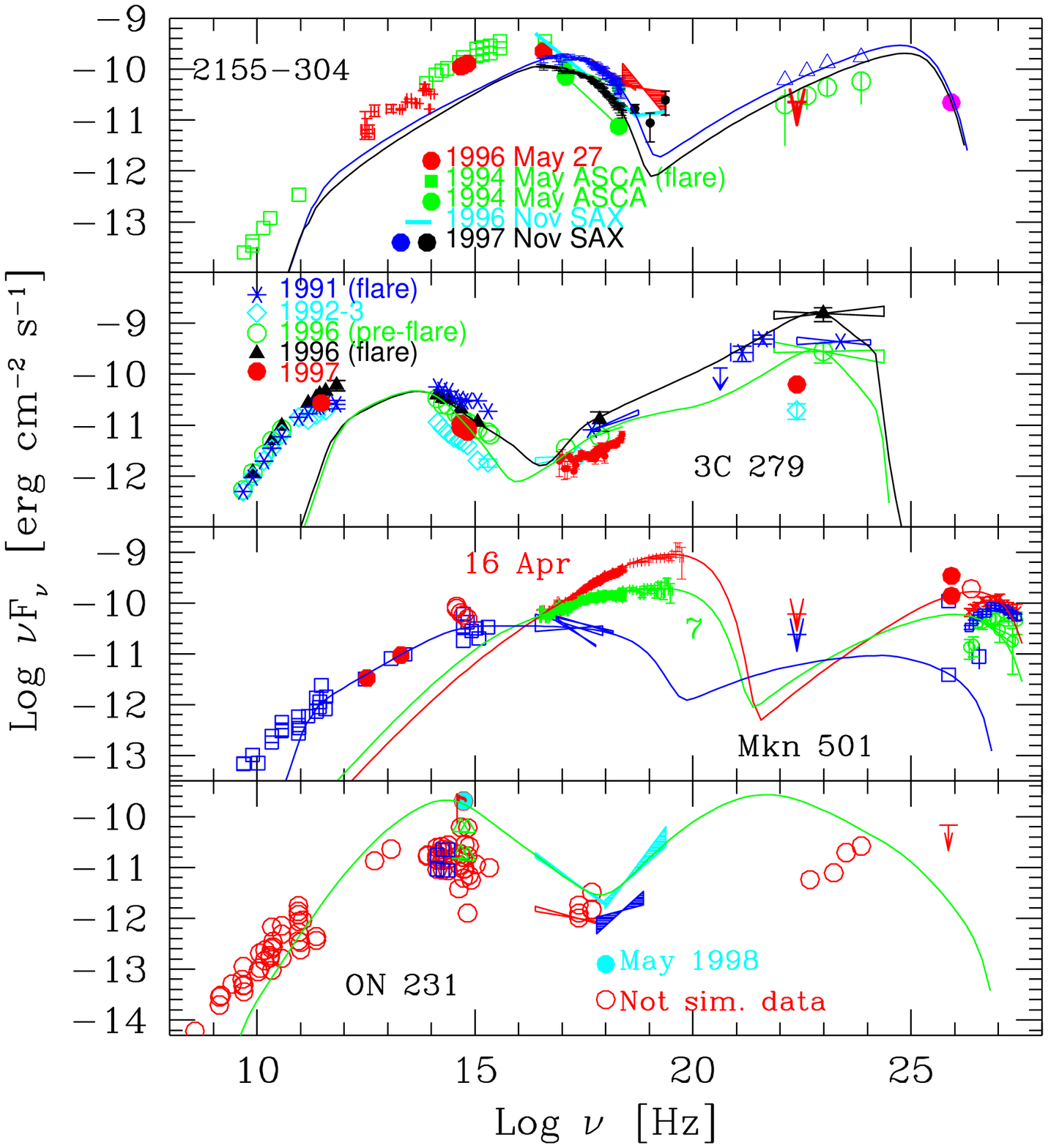,width=15truecm,height=20truecm}
\vskip -0.5 true cm
\caption[h]{Some examples of blazar SEDs, corresponding to multifrequency
observational campaigns. 
PKS 2155--304: Maraschi et al. 1998; 
3C 279: Maraschi et al. 1994, Werhle et al. 1998;
Mkn 501: Pian et al. 1998;
ON 231: Tagliaferri et al. in preparation. 
The solid lines are homogeneous synchrotron self-Compton plus external
photon models (see Ghisellini et al. 1998). }
\end{figure}

\section{The blazar SED}
Fig. 2 illustrates the overall SED of some of the best known blazars.
As can be seen, the SED is characterized by two broadly peaked components.
The first component, produced by synchrotron emission, peaks
in the mm--far IR band in powerful blazars, while it peaks at higher energies
as the total power decreases, reaching the EUV -- soft X--ray band
in HBL (high energy peaked BL Lacs).
Correspondingly, also the second more energetic component peaks at 
higher and higher energies as the total power decreases.
The correlated variability strongly suggests that 
this component is produced by inverse Compton emission by the same
population of electrons emitting the synchrotron radiation.
Moreover, this Compton component becomes increasingly dominant 
(with respect to the synchrotron one) as the total observed power increases.

\section{Unifying blazars}

The trends just mentioned have been discussed in detail by 
Fossati et al. (1998), who considered three complete samples of blazars, 
and by Ghisellini et al. (1998), who studied all blazars detected by 
EGRET for which we have some $\gamma$--ray spectral information.
In the latter paper we have modeled all sources with a homogeneous
synchrotron and inverse Compton model, taking into account, for the scattering
process, both the locally produced synchrotron photons and photons
produced externally to the jet.
The modeling allows to derive the intrinsic parameters of the sources
such as the size, the magnetic field, the energy $\gamma_{peak}m_ec^2$ of the 
electrons emitting at the peaks, the particle density and so on.
The fitted spectra are constructed through a literature search,
and are rarely simultaneous.
This implies that the fitting parameters found may be not completely
correct for individual sources, but are representative on statistical grounds.

One of the most important results of this study is the striking
correlation found between $\gamma_{peak}$ and the total (magnetic plus
radiative) energy density of the emitting region, as shown in Fig. 3.
Furthermore, $\gamma_{peak}$ also correlates with the observed power,
the amount of external seed photons used for the scattering process 
and the ratio between the Compton and the synchrotron luminosity.
The smaller $\gamma_{peak}$, the larger the energy densities and the total
power, dominated by the Compton luminosity.

Different blazar subclasses then form a well defined sequence:
\begin{itemize}
\item
powerful quasars, both polarized and not, are characterized by 
small values of $\gamma_{peak}$, so that their synchrotron spectrum
peaks in the mm-far IR band and their Compton spectrum in the MeV band.
The Compton component is stronger than the synchrotron one,
and the contribution of photons produced externally to the jet 
to the scattering process is more important than the synchrotron one. 

\item 
At the other extreme, HBL have large values of $\gamma_{peak}$,
and consequently their spectra peak in the soft-X--ray and in the 
GeV--TeV bands.
The Compton emission is roughly as powerful as the synchrotron one.
The contribution of externally produced photons is almost negligible 
(even if photons produced in regions of the jet adjacent to 
the $\gamma$--ray emitting site may be important, 
as in the case of case of Mkn 501, Pian et al. 1998).
\end{itemize}

\begin{figure}
\vskip -1 true cm
\psfig{file=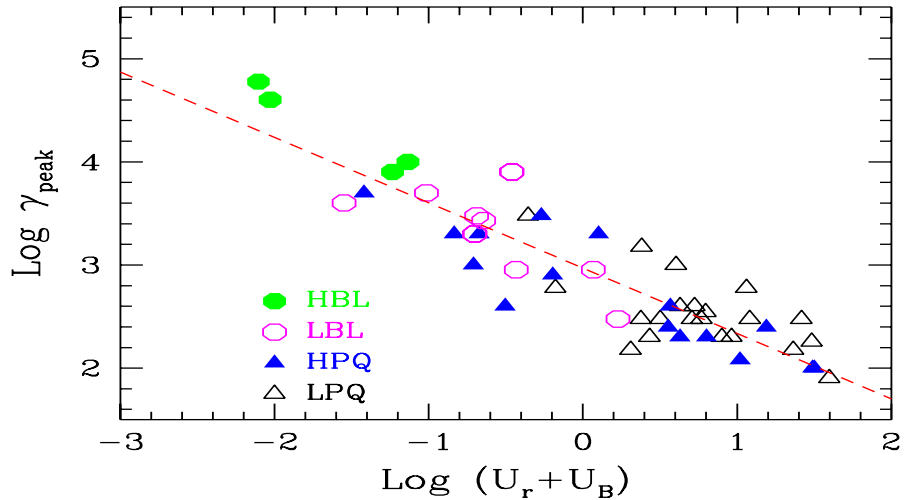,width=14truecm,height=10truecm}
\vskip -2.5 true cm
\caption[h]{The correlation between the random Lorentz 
factor of the electron emitting at the peak of the synchrotron
and the Compton spectra and the comoving energy density
(magnetic plus radiative).
Different symbols refer to subclasses of blazars, as labeled.}
\end{figure}

A tentative interpretation of these trends is that in all sources
there is a competition between the acceleration and the cooling
processes, which determines the relevant value of the electron
energy, i.e. $\gamma_{peak}$.
The particular form of the found correlations between $\gamma_{peak}$
and the (comoving) energy density $U$, $\gamma_{peak}\propto U^{-0.6}$,
implies that the radiative cooling rate 
($\dot\gamma\propto \gamma^2U$) at $\gamma_{peak}$
is almost the same for all sources.
This suggests the existence of a `universal' acceleration process
(i.e. the same in all sources, and independent of $\gamma$ and $U$)
which accelerates the electrons up to the energy where their gains
balance their losses.

%

We could then explain why only in lineless and
relatively weak BL Lacs the electrons can attain TeV energies:
in these sources the energy densities are relatively small,
and correspondingly the balance between heating and cooling rates is 
achieved at a larger $\gamma_{peak}$.
On the contrary in powerful blazars with significant emission line luminosity
the radiation energy density is large, and the balance between 
heating and cooling is achieved at a smaller $\gamma_{peak}$.
In these sources the Compton cooling is enhanced (because of greater radiation
energy densities), and this is why the Compton to synchrotron 
luminosity ratio is larger.

\subsection{How to falsify the proposed scenario}

In the proposed picture blazars lye along a sequence:
from powerful flat spectrum radio sources to less powerful lineless BL Lacs.
The electron cooling rate is the key parameter ruling the value
of $\gamma_{peak}$ and therefore their overall SED.

This scenario could then be falsified if
there exist powerful blazars with relatively strong emission lines and
with synchrotron and Compton peaks located at high energies 
(in the soft X--rays and in the GeV-TeV band).

Instead low luminosity blazars with a SED peaking at low energies
(far IR and MeV--GeV) would not necessary contradict our picture,
since they can be slightly misaligned powerful blazars.

\section{The power of blazar jets}

Jets must power the extended radio lobes of radiogalaxies and
quasars, which contain a huge amount of energy.
By estimating it with equipartition arguments, and dividing it by the age
of the source, one obtains the average power that lobes needed to
be formed and to grow (Rawlings and Saunders 1991).
This can amount to $10^{46}$--$10^{47}$ erg s$^{-1}$.
Jets have to carry this amount of power in the form of bulk kinetic flow 
($L_k$) and/or Poynting vector ($L_B$).
These powers can be estimated at any location in the jet 
knowing the corresponding cross sectional jet area ($\pi R^2$), 
the particle density ($n=n^\prime/\Gamma$, where
$n^\prime$ is the density in the comoving frame), 
the bulk Lorentz factor ($\Gamma=1/\sqrt{1-\beta^2}$), 
the magnetic field ($B$, as measured in the comoving frame), 
the matter content (electron--positron pairs or electron--protons) 
and the average internal energy of the leptons 
($<\gamma> m_ec^2$, assuming that the protons are cold).
We then have (Celotti \& Fabian 1993, Ghisellini \& Celotti 1998):

\begin{equation}
L_k\, =\, \pi R^2 \Gamma^2\beta c^3 n^\prime (<\gamma> m_e + m_p)
\end{equation}
\begin{equation}
L_B\, =\,  {1\over 8}\, R^2 \Gamma^2\beta c B^2 
\end{equation}
where we have assumed a proton--electron plasma.
The observed synchrotron luminosity of a blob of volume $(4\pi/3)R^3$,
observed at an angle $1/\Gamma$ is
\begin{equation}
L_{s,obs} \, =\, {4\pi R^3\over 3}
\Gamma^4 \int n^\prime(\gamma)\dot\gamma_s m_ec^2d\gamma\, 
= \, {2 R^3 \over 9} \Gamma^4 \sigma_T c n^\prime B^2 <\gamma^2>
\end{equation}
where $<\gamma^2>$ is averaged over the emitting particle distribution.
By substituting the particle density derived by this equation into Eq. (1),
we have
\begin{equation}
L_k\, =\, {9\pi m_ec^2 \over 2\sigma_T}\, 
{L_{s,obs} \over R \Gamma^2 B^2}
\, {<\gamma> + m_p/m_e \over <\gamma^2>}
\end{equation}
We see that $L_k \propto (B\Gamma)^{-2}$, while $L_B\propto (B\Gamma)^2$:
therefore $L_{jet}\equiv L_k+L_B$ is minimized for some value of 
$B\Gamma$, which corresponds
to equipartition between particle and magnetic energy densities
(Ghisellini \& Celotti, 1998; see also Celotti \& Ghisellini, this volume).
In this case the $observed$ (at a viewing angle $\sim 1/\Gamma$)
synchrotron emission is maximized.
In other words, the most economic way for a blazar to emit the 
observed synchrotron power is to be in equipartition.
Fig. 4 shows that blazars can really be in equipartition: for this
figure we have used the parameters ($B$ field, particle distribution,
dimension) derived by fitting the spectra of the EGRET blazars
(Ghisellini et al. 1998).
One can see that for $\Gamma\sim 10$--15 (which is the value used for the 
fits) we have the minimum energetic requirement.
In this figure we also plot the radiative total luminosity $L_r$,
which is the power emitted in the rest frame multiplied
by $\Gamma^2$, i.e. it is the power received over the entire solid angle
(it is $not$ the power inferred from the flux multiplied by $4\pi d^2$,
where $d$ is the luminosity distance).
If the radiative power originates from the conversion of bulk into random 
kinetic energy, and in steady state conditions, $L_r$ cannot exceed $L_k$.
However, $L_r$ approaches $L_k$ in the case of Mkn 501,
suggesting that extreme flaring states (the power of Mkn 501
during the 1997 flare was 20--fold the power in the quiescent state) 
may correspond to more efficient bulk to random energy conversion.

Another interesting consequence is that 
{\it there is a limit to $\gamma_{peak}$}, at least in steady state sources:
the higher $\gamma_{peak}$, the faster the cooling rate, and 
the greater the radiated luminosity.
But the latter cannot exceed the power in bulk motion,
and thus sets an upper limit to $\gamma_{peak}$ (which might have
been almost reached by Mkn 501 during its flare).

\begin{figure}
\vskip -1.4 true cm
\psfig{file=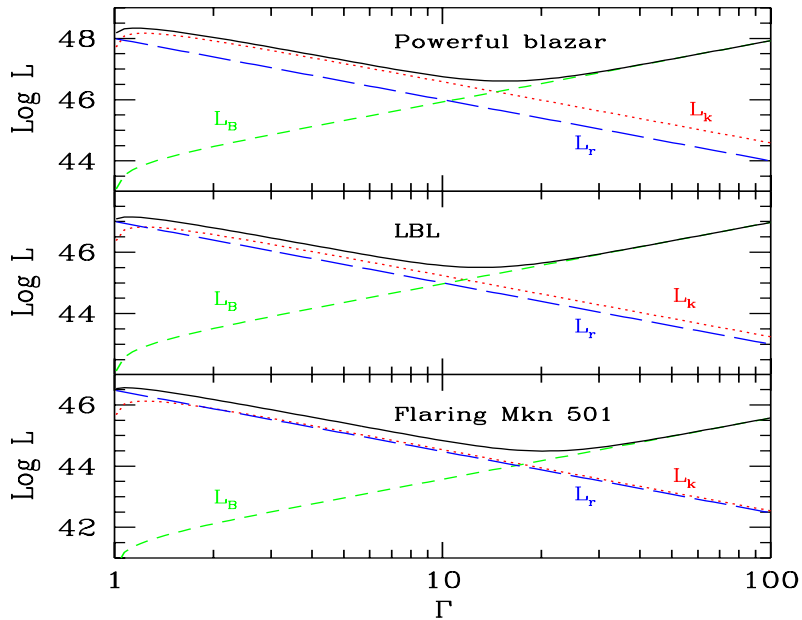,width=13truecm,height=10truecm}
\vskip -1.2 true cm
\caption[h]{The power in the form of bulk kinetic energy ($L_k$),
Poynting vector ($L_B$) and radiation ($L_r$) as a function of the
bulk Lorentz factor $\Gamma$ for a powerful flat spectrum quasar,
a low energy peak BL Lac (LBL) and for the extreme BL Lac Mkn 501
assuming the spectrum of the 1997 flare. From Ghisellini \& Celotti, 1998}
\end{figure}

\section{Jet content}
We know the power required by the extended radio lobes to exist.
We can equate this power to the power in bulk kinetic motion of the 
jet plasma, and infer if the jet is composed by electron--positron 
pairs or by an electron--proton plasma.
The crucial quantity we need to know is the particle density in the jet.
We can estimate it through the synchrotron emission of the relativistic 
particles, integrating over the particle distribution $n(\gamma)$. 
Unfortunately, this integral depends on the low energy end 
of the distribution, $\gamma_{min}$:
at these energies the electrons 
emit self--absorbed synchrotron radiation and are therefore unobservable.
The other quantity required is the dimension of the region emitting this  
radiation, which can be directly obtained through VLBI observations.
In this way Celotti \& Fabian (1993) derived two possible solutions:

i) $\gamma_{min}\sim 1$, and the jet is $e^\pm$ dominated;

ii) $\gamma_{min}\sim 30$--100, and the jet is made by electrons and protons.

Note that pairs could not be formed at the base of the jet, since they 
would not survive the strong annihilation implied by their large density.
Furthermore, for what discussed in \S 2, it is problematic
to use the $\gamma$--ray radiation to create them along the jet, 
since an inevitable by--product of this would be
an excessive X--ray radiation.
For these reasons I prefer solution ii) (preferred also by Celotti and Fabian). 
For reviews on the problem of the matter content of jet, see Celotti (1997
and 1998).

\section{Conclusions}
\noindent
\begin{itemize}
\item Strong and variable $\gamma$--ray emission implies that the high 
energy radiation is beamed, and is produced far away from the black hole
and the accretion disk. 
This points to an energy transport mechanism which is dissipationless
up to the $\gamma$--ray production site.

\item Electron--positron pairs cannot be created by absorbing a significant
amount of $\gamma$--rays. This would produce too many X--rays.

\item Variability is complex, but the correlations observed in different
bands suggest a single population of electrons, emitting by the synchrotron
and the inverse Compton processes.

\item Blazars form a sequence. Their relevant electron energy
correlates with the comoving energy density, the observed power, the
Compton to synchrotron power ratio and the amount of external
seed photons scattered at high energies, which can be
identified with photons from the broad emission line region.
Low luminosity BL Lacs are the best candidates to be TeV emitters,
powerful blazars are strong MeV sources.

\item Blazars are close to equipartition between the 
power in bulk kinetic motion 
and the Poynting flux. This is the most economic way to produce the synchrotron
emission we observe.

\end{itemize}

\acknowledgments
It is a pleasure to thank Annalisa Celotti and Laura Maraschi for
constant help.

\end{document}